\documentclass[12pt]{article}

\usepackage{scicite}

\usepackage{times}

\usepackage{amsmath}
\usepackage{physics}
\usepackage{graphicx}
\usepackage{gensymb}

\newenvironment{sciabstract}{%
\begin{quote} \bf}
{\end{quote}}

\newcounter{lastnote}

\title{Coherent Dipolar Coupling between Magnetoelastic Waves and Nitrogen Vacancy Centers}

\author
{Adi Jung,$^{1}$ Samuel Margueron,$^{2}$ Ausrine Bartasyte$^{2,3,4}$ Sayeef Salahuddin$^{1\ast}$\\
\\
\normalsize{$^{1}$Department of Electrical Engineering, University of California, Berkeley}\\
\normalsize{$^{2}$FEMTO-ST Institute, University of Franche-Comté ENSMM, 26 rue de l’Epitaphe, Besancon 25030, France}\\
\normalsize{$^{3}$Institut Universitaire de France}\\
\normalsize{$^{4}$Université Paris-Saclay, CNRS, Centre de Nanosciences et de Nanotechnologies, 91120, Palaiseau, France}\\
\\
\normalsize{$^\ast$To whom correspondence should be addressed; E-mail:  sayeef@berkeley.edu.}
}

\date{}

\begin{document} 


\baselineskip24pt


\maketitle


\begin{sciabstract}
    We experimentally demonstrate coherent Rabi oscillations of Nitrogen Vacancy (NV) centers by magnetoelastic waves. The coupling is consistent with dipolar stray field drive from spin-wave modes in a ferromagnetic film, and displays a significant improvement in Radio Frequency power efficiency relative to other methods of microwave excitation. Further, it demonstrates coherent coupling with NV centers over mm-scale distances from the microwave excitation source. By utilizing a piezoelectric-magnetostrictive heterostucture, where magnetoelastic waves can be launched by an applied voltage, a pure voltage driven coherent drive of the NV centers is achieved. This voltage driven, magnetoelastic excitation enables a new approach to couple with two level quantum states that is not reliant on long spin-wave coherence lengths.
\end{sciabstract}


\paragraph*{Introduction}

Over the past decade, hybrid quantum magnonic systems have emerged as a novel approach to enhancing quantum defect center performance and implementing integration in practical systems\cite{lachance-quirion_hybrid_2019,li_hybrid_2020,fukami_opportunities_2021}. In particular, a range of studies have been performed utilizing spin waves as an interface with the nitrogen-vacancy (NV) defect center in diamond\cite{bertelli_magnetic_2020,wolfe_off-resonant_2014}. These approaches have enabled new measurement techniques for probing magnetic films \cite{du_control_2017}, and have presented an approach for improving power efficiency of the microwave coupling by using a magnetic film as an effective amplifier \cite{andrich_long-range_2017}.

However, despite these advantages, existing approaches relying on spin wave drive of nitrogen-vacancy centers continue to suffer from various limitations. While spin waves effectively enhance the range of a microwave drive antenna, this enhancement is limited to the propagation length of spin waves, which is typically in the 10s of microns \cite{liu_long-distance_2018}. In addition, the overwhelming majority of these demonstrations have used single-crystal yttrium iron garnet (YIG) films as the magnetic media, which often requires the use of a gadolinium gallium garnet (GGG) substrate to achieve long coherence lengths. YIG is a model magnetic system for magnonics as a consequence of its high quality and low spin wave damping, but YIG thin films are highly challenging to integrate in practical devices in large part due to their growth requirements \cite{schmidt_ultra_2020}. 

Independent of the development of hybrid quantum magnonics, magnetoelastic devices have also recently emerged as a potential approach for exciting magnetic dynamics in a highly power efficient manner\cite{labanowski_power_2016,dreher_surface_2012,chowdhury_nondegenerate_2017}. Elastic waves are capable of propagating over much longer length scales than spin waves, and are capable of coherent conversion to spin wave modes when traveling through magnetic films\cite{hioki_coherent_2022,li_spin_2017}. Acoustically driven ferromagnetic resonance has already been shown to couple off-resonantly to the nitrogen-vacancy center, but this mechanism involves incoherent scattering of spin wave modes, making it challenging to use for quantum information and magnetometry\cite{wolfe_off-resonant_2014,labanowski_voltage-driven_2018}. 

In this paper, we demonstrate that dipolar stray field coupling of magnetoelastic waves in nickel thin films to NV centers is phase-coherent. Furthermore, we show that this excitation displays strong Optically Detected Magnetic Resonance (ODMR) contrast at orders of magnitude lower microwave input power relative to existing approaches, and is expected to be generalizable to a wide range of magnetostrictive materials. 

\begin{figure}
\includegraphics[width=\linewidth]{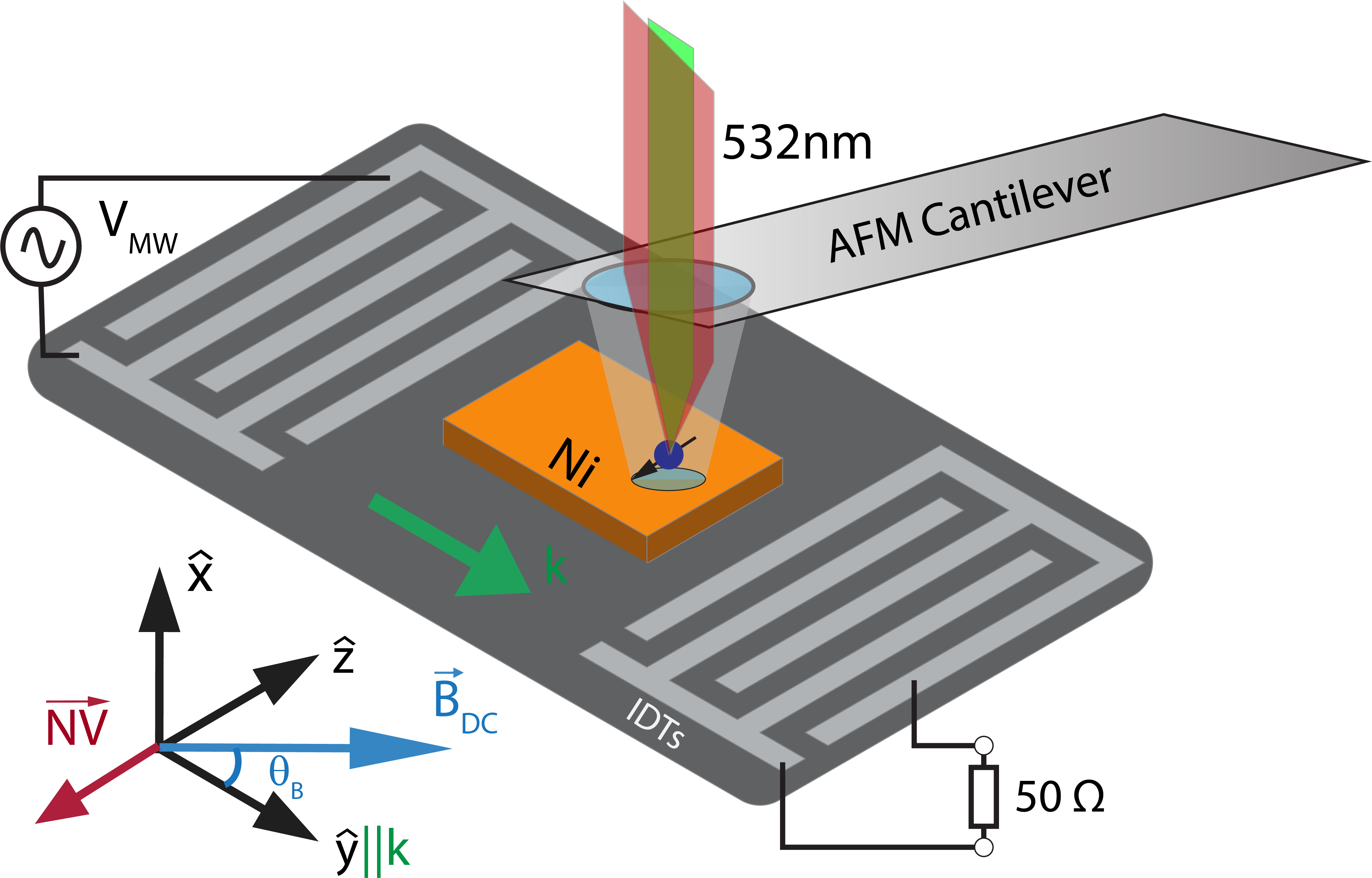}
\caption{Diagram of experimental configuration. The AFM tip was spaced at distance of 150 nm from the Ni film surface, and centered within the acoustic path 10 $\mu$m away from the leading edge of the ferromagnetic film. Piezoelectric transducers were designed to have input impedances of 50 $\Omega$, and the receiver was loaded with a matching impedance to reduce acoustic reflections. }
\end{figure}

\begin{figure}
\includegraphics[width=\linewidth]{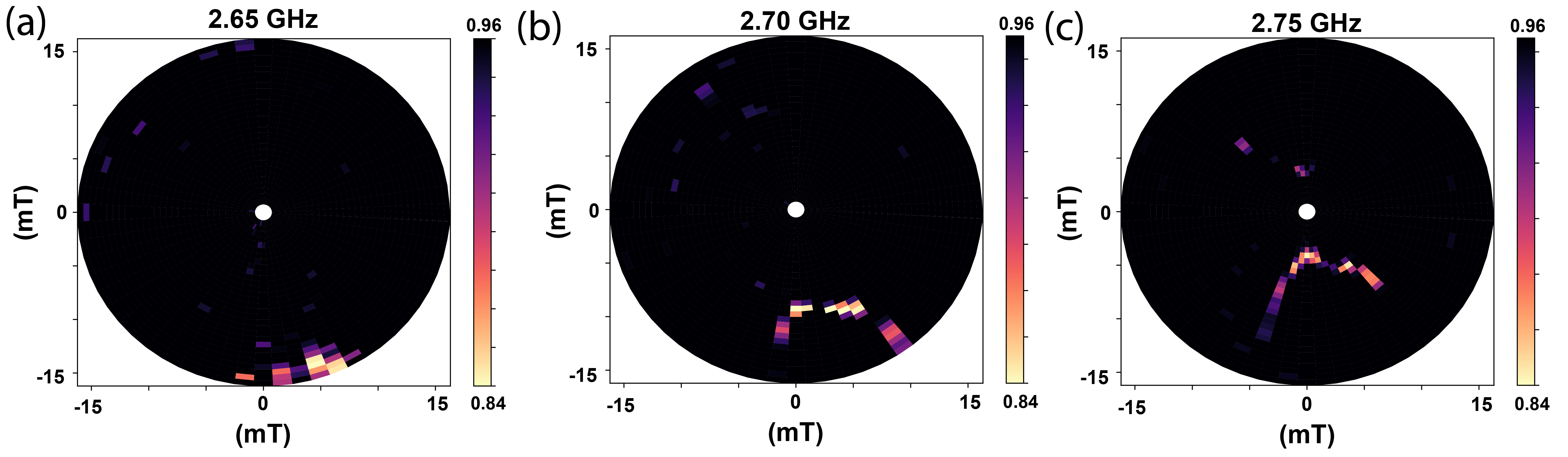}
\caption{ODMR contrasts measured at various microwave frequencies under radially swept magnetic bias fields with an input RF power of -18 dBm. The strongest ODMR coupling is observed at bias field angles of 270\degree, consistent with prior theory and experiment, as discussed in literature \cite{jung_coupling_2024}. The frequency dependence of the pattern displays a qualitative shift consistent with the Zeeman effect; as frequency approaches the NV-center transition frequency of 2.87 GHz, the magnetic bias field at which ODMR occurs is reduced. }
\end{figure}

\begin{figure}
\includegraphics[width=\linewidth]{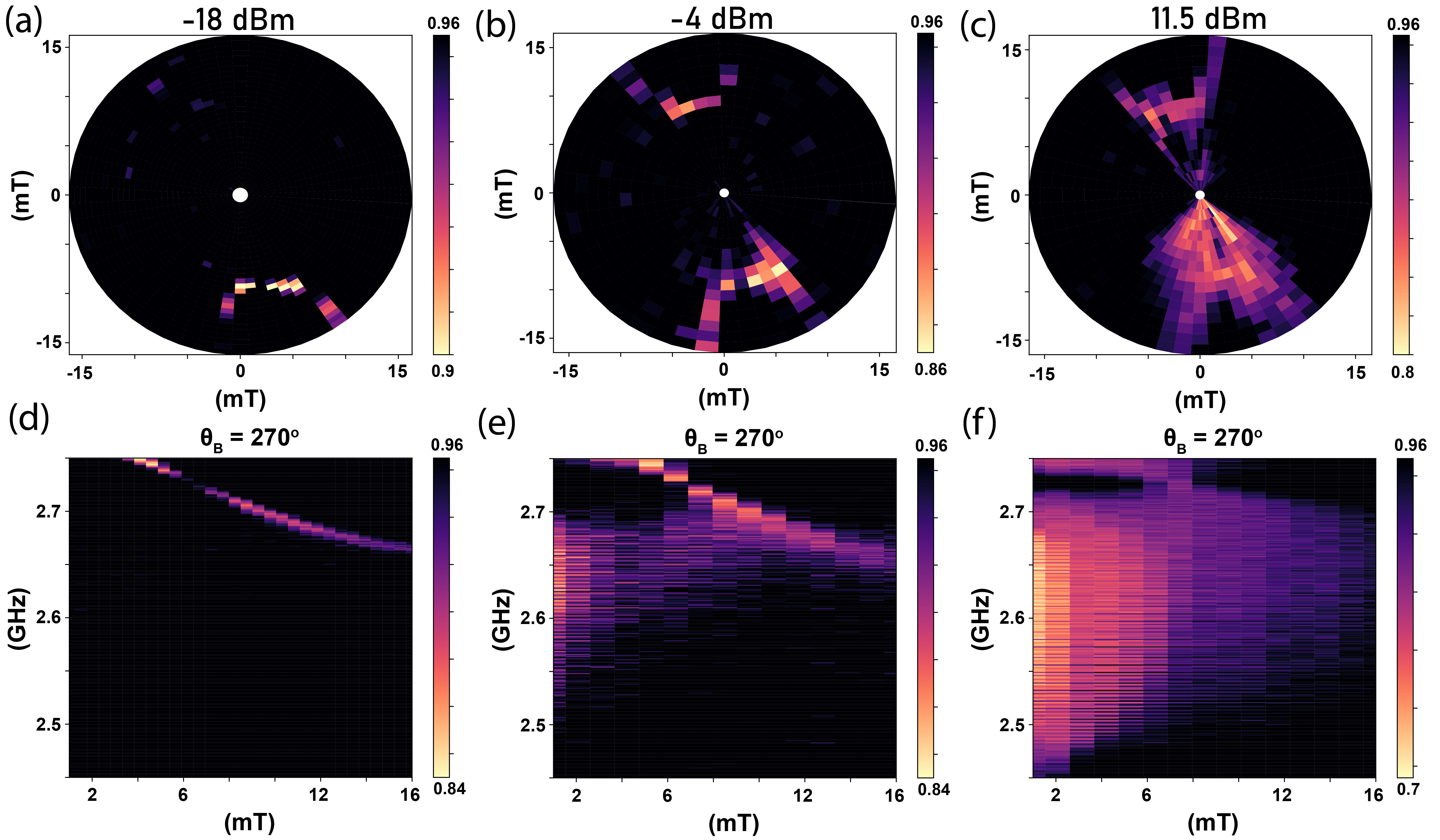}
\caption{(a-c) ODMR contrast measured at 2.70 GHz, under radially swept magnetic bias field at varying RF power levels. As RF power increases, ODMR linewidth broadening is observed, and a weaker response at 90$\degree$ becomes visible. This is attributed to a lower power surface acoustic wave reflection from the receiving acoustic transducer. (d-f) ODMR spectra measured under bias field angle of 270$\degree$ at varying power. At low power, a feature qualitatively consistent with a two-level transition appears. As power is increased, a large broadband feature consistent with off-resonant FMR coupling appears \cite{wolfe_off-resonant_2014}\cite{labanowski_voltage-driven_2018}. At high power, the frequency dependence of this off-resonant ODMR feature is consistent with the transmission band of the acoustic transducers, providing further evidence that this feature coincides with high delivered acoustic power.  }
\end{figure}

\paragraph*{Results}
Plotted in Figure 2 is the observed magnetic bias field dependence of the NV center fluorescence contrast. As previously shown via theory and experiment \cite{yu_chiral_2019} \cite{jung_coupling_2024}, the fluorescence contrast due to a magnetoelastic wave is, for this NV orientation, expected to occur when the magnetic bias field is at a 270$\degree$ angle to the surface acoustic wave propagation direction. The measured fluorescence contrast seen in Figure 2 displays this property, as well as a frequency dependent shift consistent with the Zeeman effect. The slight rotation offset of approximately 5$\degree$ is due to hand mounting error that resulted in misalignment between the device and the external electromagnet. The resulting fluorescence contrast varied depending on magnetic bias field and drive frequency, but was typically between 10\% and 15\% for an input power of 16 $\mu$W.

Further confirmation that the observed fluorescence contrast corresponded to a dipolar drive was provided by the ODMR spectrum characterized at various RF power levels in Figure 3. The ODMR spectrum in Figure 3d clearly displays a feature consistent with a two-level transition. The nonlinear relationship between frequency and applied bias field is attributed to alignment error of the NV-center axis. In this measurement, the NV center axis was subject to misalignment due to the nature of the diamond cantilever tip used; in typical NV-center studies the magnetic bias field would have been adjusted parallel to the NV-axis to compensate for this misalignment. However, for studies of magnetoelastic wave drive, the relative orientation of the magnetic bias field and acoustic propagation is critical, and this bias field adjustment would have acted as a confounding variable. As the relationship between the NV-center and magnetic field has already been studied comprehensively, we elected to align the bias field according to the SAW propagation axis instead. This leads to the measured quantitative discrepancy between the observed ODMR spectrum and the Zeeman effect, and is qualitatively consistent with previous measurements of systems with misaligned bias field \cite{jeong_understanding_2017}.

Multiple effects not visible under low power excitation can be observed as drive power is increased. In addition to the expected linewidth broadening of the ODMR resonance, a broadband fluorescence contrast feature appears, consistent with prior measurements of ferromagnetic resonance-coupled NV centers \cite{wolfe_off-resonant_2014} \cite{labanowski_voltage-driven_2018}. An additional weaker fluorescence contrast feature also appears at a 90$\degree$  bias field angle to surface acoustic wave propagation. This feature appears similar to the primary 270$\degree$  feature, but would be expected to appear if the acoustic propagation direction was reversed. It is well-established that in interdigitated transducers such as those used in this work, multi-transit and reflected acoustic wave signals can be observed electrically, and that first order reflections waves can have meaningful amplitude relative to the primary pulse. As such, these weaker fluorescence contrast features are attributed to acoustic reflections, a claim further corroborated by Rabi oscillation measurement discussed in Figure 4. Also visible in Figure 3f is a distinct frequency-dependence of the off-resonant FMR-coupling feature, wherein the fluorescence contrast is maximized at 2.6 GHz, and heavily reduced around 2.45 GHz and 2.725 GHz. This feature is unrelated to magnetic properties, rather, the frequency band over which the fluorescence is maximized corresponds to the transmission bandwidth of the acoustic transducers. The IDT transmission band also explains the apparent break in the ODMR spectrum seen in Figure 3d, as negligible acoustic power is transmitted at that frequency.

Key to the demonstration of phase-coherent coupling was the identification of experimental conditions under which the dipolar coupling dominated over off-resonant mechanisms related to acoustically driven ferromagnetic resonance \cite{labanowski_voltage-driven_2018}. Based on the results of Figures 2 and 3, we selected a magnetic bias field of 10 mT at 270$\degree$ and an RF input power of -18 dBm. Under these bias conditions, the ODMR resonance occurred centered at a drive frequency of 2.679 GHz. The fluorescence contrast measured under these bias and drive conditions for varying RF pulse durations is provided in Figure 4a, along with a Rabi oscillation fitting with a frequency of 3.17 MHz. The fit is qualitatively sound except for a region corresponding to pulse durations between 0.7 and 0.9$\mu$s. As mentioned briefly above, this can indirectly be explained by SAW reflections, as the average SAW velocity in this device, accounting for relative length of metallized and unmetallized components, is approximately 3926 m/s \cite{soluch_surface_2005}. For the geometry of the device and NV tip position used here, consisting of a 2mm long acoustic delayline and a 0.5mm long Ni film, with tip positioned approximately at the leading edge of the Ni film, reflected acoustic waves from the receiver are expected to begin to overlap with the primary pulse for pulse durations greater than 0.64 $\mu$s, and are hypothesized to decohere the primary acoustic excitation.

Following this, we compare the effective power efficiency of ODMR contrast measured here to that characterized across a range of techniques within the literature, as shown in Figure 4b. While this is an approximate comparison, as ODMR contrast varies depending on source proximity to the NV center, which is challenging to quantify, most practical laboratory measurements using antennas require several hundred milliwatts of RF power to exceed an ODMR contrast of 10 percent \cite{kim_cmos-integrated_2019}, \cite{sasaki_broadband_2016}. A more direct comparison can be performed to spin-wave based RF-drive \cite{andrich_long-range_2017}, where the NV centers were positioned in direct contact to a ferromagnetic film's surface. Spin wave-based excitation of NV centers is being actively explored as an approach to reducing RF power consumption for NV center drive. However, this work suggests that not only is a magnetoelastic drive a more flexible approach to this excitation, as it is compatible with magnetic materials with short spin-coherence lengths, but it also significantly outperforms direct spin wave excitation in terms of power consumption. Under slightly reduced input power, magnetoelastic ODMR contrast was an order of magnitude larger than a comparable RF power spin-wave drive, suggesting that magnetoelastic drive is a highly promising approach toward practical small-scale integration of NV-centers.  

\begin{figure}
\includegraphics[width=\linewidth]{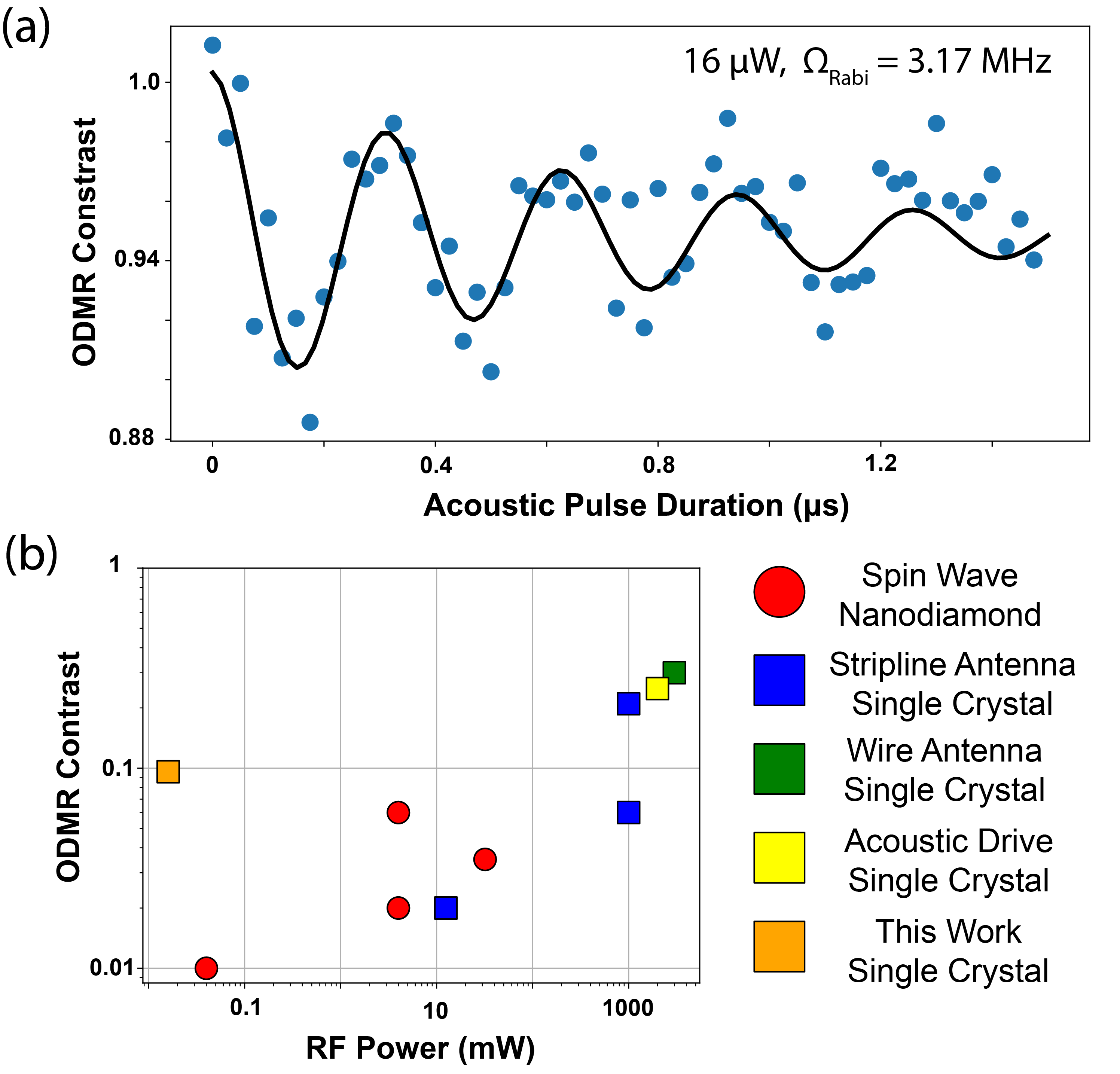}
\caption{(a) Fluorescence contrast under variation of applied microwave pulse duration. Rabi oscillations remain clearly coherent for pulse durations below 0.6 $\mu$s. The deviation from the Rabi fit between 0.7 and 0.9 $\mu$s can be explained by SAW reflections. For the substrate and delayline geometry used, the surface acoustic wave would be expected to overlap with its reflection at the location of the NV center for pulses longer than approximately 0.64 $\mu$s, leading to potential decoherence. (b) Comparing RF power consumption to various sources in literature \cite{andrich_long-range_2017} \cite{jeong_understanding_2017} \cite{kim_cmos-integrated_2019} \cite{macquarrie_coherent_2014} \cite{opaluch_optimized_2021} \cite{sasaki_broadband_2016}, we see that magnetoelastic drive of NV centers displays multiple orders of magnitude superior power efficiency to a a wide range of alternative approaches of microwave drive.}
\end{figure}

\paragraph*{Conclusions}
In short, we have demonstrated a coherent coupling between magnetoelastic waves and the Nitrogen-vacancy center in diamond. The coupling is demonstrated to be mediated by the dipolar fields of magnetoelastic spin dynamics, and displays an exceptionally efficient power coupling over millimeter length scales. In comparison to results previously demonstrated using YIG spin waves, this magnetoelastic approach achieves superior power efficiency unrestricted by the propagation losses and coherence length of spin waves. The approach is also compatible with existing MEMS fabrication technology, and under appropriate design constraints, is potentially applicable across all use cases of NV centers, ranging from magnetometry to quantum information processing. 

In addition, while these results serve as a proof-of-concept, the design parameters chosen in this work have significant room for improvement. According to linear magnetoelastic theory, longitudinal acoustic waves are not expected to induce magnetic dynamics under magnetic bias conditions where to the NV center is maximized\cite{kamra_coherent_2015,kus_chiral_2022}. The Rayleigh mode wave excited here is dominated by a longitudinal component, with a shear component which is a factor of 10 weaker \cite{jung_double-peaked_2021}. However, a shear-mode dominated surface acoustic wave device fabricated on a substrate such as $LiTaO_3$ \cite{takai_high-performance_2017}, would likely  display a comparable ODMR contrast at lower input power. In conjunction with efficient piezoelectric transducers and a more magnetostrictive magnetic material, power efficiency can plausibly be improved by 2-3 orders of magnitude beyond the results displayed here.  

\paragraph*{Acknowledgements} This work was supported by the NSF TANMS center within National Science Foundation Cooperative Agreement EEC-1160504. This work was also supported by the EIPHI Graduate School contract ANR 17 EURE 0002, ANR project MAXSAW ANR-20-CE24-0025, the French RENATECH network and its FEMTO-ST technological facility. The authors are grateful for the support of and discussions with Dr. Jan Rhensius and the QZabre AG team.

\paragraph*{Methods}
The magnetoelastic device in this study consists of a piezoelectric substrate (128$\degree$ Y-cut, X-propagating LiNbO$_3$) on which surface acoustic wave (SAW) transducers are fabricated (Al, 80 nm thick), with a center frequency of approximately 2.6 GHz. A polycrystalline magnetic film (Ni, 20 nm thick) was evaporated and lift-off patterned between the IDTs. NV center ODMR and Rabi oscillations were both measured using a commercial NV magnetometry microscope (Quantum Scanning Microscope, QZabre AG), which positions a spatially scanning cantilever containing a single NV center at an adjustable spacing from the magnetic film, as shown in Fig 1. All data was collected with the tip at a distance of 150 nm from the Ni surface, positioned approximately 10 $\mu$m from the leading edge of the Ni film and centered within the acoustic wave propagation path. The transmitter transducer was driven with a microwave power level of -18 dBm, and the receiver was fed into a 50 $\Omega$ load. The transducer spacing was 2 mm, and the distance between the transmitter to AFM tip characterization position was 775 $\mu$m. Applied green laser power was set to 2 mW.
Continuous wave ODMR spectra were collected over frequencies ranging from  of 2.45 GHz to 2.75 GHz, containing the central band of the interdigitated transducer filter function which ranges from 2.4 GHz to 2.7 GHz. Owing to this transducer bandwidth limit, meaningful data could not be acquired at frequencies significantly outside this range. The ODMR spectra were collected while magnetic field was swept in both magnitude and direction in the plane of the magnetic film. To ensure a predictable, fully saturated initial magnetization state, prior to data collection at each angle a magnetic field of 40mT was applied to saturate the magnetic film before the amplitude sweep.

\bibliography{06-08-2024}

\bibliographystyle{Science}

\clearpage

\end{document}